\documentclass[useAMS,usenatbib]{mn2e}
\usepackage{natbib}
\usepackage{graphicx}

\title[Numerical Modelling of Palomar 14]{Direct N-body simulations of globular
clusters: (I) Palomar 14}

\author[Hasani Zonoozi et al.]
{Akram Hasani Zonoozi$^{1,2}$\thanks{
E-mail:  \mbox{a.hasani@iasbs.ac.ir} (AHZ)   %;
\mbox{akuepper@astro.uni-bonn.de} (AK);
\mbox{h.baumgardt@uq.edu.au}(HB);
 \mbox{haghi@iasbs.ac.ir}(HH);
 \mbox{pavel@astro.uni-bonn.de} (PK);\mbox{mhilker@eso.org} (MH)
 }
, Andreas H.W. K\"{u}pper$^{2,3}$, Holger Baumgardt$^{2,4}$,
\newauthor
Hosein Haghi$^{1,2}$, Pavel Kroupa$^2$, Michael Hilker$^5$\\
$^{1}$Department of Physics, Institute for Advanced Studies in Basic Sciences (IASBS), P.O. Box 11365-9161, Zanjan, Iran\\
$^{2}$Argelander Institute f\"ur Astronomie (AIfA), Auf dem H\"ugel 71, 53121 Bonn, Germany\\
$^{3}$European Southern Observatory, Alonso de Cordova 3107, Vitacura, Santiago, Chile\\
$^{4}$University of Queensland, School of Mathematics and Physics, Brisbane, QLD 4072, Australia\\
$^{5}$European Southern Observatory, Garching b. M\"{u}nchen,
Germany \\}

\begin{document}

\date{Accepted \ldots. Received \ldots; in original form \ldots}

\pagerange{\pageref{firstpage}--\pageref{lastpage}} \pubyear{2010}

\maketitle

\label{firstpage}

\maketitle

\begin{abstract}
We present the first ever direct $N$-body computations of an old
Milky Way globular cluster over its entire life time on a
star-by-star basis. Using recent GPU hardware at Bonn University, we
have performed a comprehensive set of $N$-body calculations to model
the distant outer halo globular cluster Palomar 14 (Pal 14). Pal 14
is unusual in that its mean density is about ten times smaller than
that in the solar neighborhood. Its large radius as well as its
low-mass make it possible to simulate Pal 14 on a star-by-star
basis. By varying the initial conditions we aim at finding an
initial $N$-body model which reproduces the observational data best
in terms of its basic parameters, i.e. half-light radius, mass and
velocity dispersion. We furthermore focus on reproducing the stellar
mass function slope of Pal 14 which was found to be significantly
shallower than in most globular clusters.

While some of our models can reproduce Pal 14's basic parameters
reasonably well, we find that dynamical mass segregation alone
cannot explain the mass function slope of Pal 14 when starting from
the canonical Kroupa initial mass function (IMF). In order to seek
for an explanation for this discrepancy, we compute additional
initial models with varying degrees of primordial mass segregation
as well as with a flattened IMF. The necessary degree of primordial
mass segregation turns out to be very high, though, such that we
prefer the latter hypothesis which we discuss in detail. This
modelling has shown that the initial conditions of Pal 14 after gas
expulsion must have been a half-mass radius of about 20 pc, a mass
of about 50000 M$_{\odot}$, and possibly some mass segregation or an
already established non-canonical IMF depleted in low-mass stars.
Such conditions might be obtained by a violent early gas-expulsion
phase from an embedded cluster born with mass segregation. Only at
large Galactocentric radii are clusters likely to survive as bound
entities the destructive gas-expulsion process we seem to have
uncovered for Pal 14.

In addition we compute a model with a 5\% primordial binary fraction
to test if such a population has an effect on the cluster's
evolution. We see no significant effect, though, and moreover find
that the binary fraction of Pal 14 stays almost the same and gives
the final fraction over its entire life time due to the cluster's
extremely low density. Low-density, halo globular clusters might
therefore be good targets to test primordial binary fractions of
globular clusters.

\end{abstract}


\section{Introduction}

The initial mass function (IMF) of stars is an important quantity
for astrophysics. Since the properties and evolution of a star are
closely related to its mass, the IMF is a link between stellar
evolution and the evolution of stellar systems. It influences the
chemical evolution of galaxies and the dynamical evolution of star
clusters. The IMF is often expressed as a power-law function
($\frac{dN}{dm}\propto m^{-\alpha}$). The traditional IMF that was
proposed by Salpeter (1955) from analyzing stars in the solar
neighborhood with masses between $0.4 \mbox{M}_{\odot}$ and $10
\mbox{M}_{\odot}$, has $\alpha=2.3$, implying that high mass stars
are rare compared to low mass stars. Later advances showed that the
IMF slope flattens to $\alpha=1.3$ for stars in the mass range $0.5
\mbox{M}_{\odot}-0.08 \mbox{M}_{\odot}$ while it remains Salpeter's
for massive stars \citep{kro01}.

Since stars in star clusters have similar metallicities, distances
and ages, they are the best objects to study the stellar mass
function. However, the mass function of stars in clusters evolves
through stellar and dynamical evolution and one has to understand
how these processes have affected the mass function of stars before
one can extract the IMF from the observed mass function (Baumgardt
\& Makino 2003). For example, as a result of energy equipartition,
the heavier stars tend to move toward the cluster centre while
lighter stars tend to move further away from the cluster centre.
Since observations of stars in globular clusters usually are
radially limited, the effect of mass segregation has to be
considered when deducing the global mass function from the observed
local one. Moreover, as a result of preferential loss of low-mass
stars, a cluster's mass-to-light ratio evolves differently from that
expected from pure stellar evolution, such that dynamical evolution
leads to a decrease of the mass-to-light ratio
\citep{bau03,krui09,and09}.

While the observed mass function shows local variations, e.g. as a
consequence of dynamical mass segregation, the IMF is usually
assumed to have no spatial variation throughout a cluster. However
star formation simulations indicate that this assumption could be
invalid (e.g., Tan et al 2006; Krumholz et al 2009). There is also
some observational evidence for deviations from the standard initial
mass function, at low and high mass ends, in many star clusters,
which might be the result of dynamical evolution (e.g., see
Elmegreen 2004 for review).

Some of the ideas that have been proposed to explain a shallowness
of the slope at the high mass end include primordial mass
segregation of stars in the cluster (e.g., Vesperini \& Heggie 1997;
Kroupa 2002; Mouri \& Taniguchi 2002). But so far it is not clear if
star clusters were primordially mass segregated from the star
formation epoch or obtained mass segregation via dynamical evolution
at a later time. Especially for some young star clusters it has been
shown that the observed mass segregation cannot be explained
dynamically from an initially unsegregated system (Bonnell \& Davies
1998). The early mass loss due to stellar evolution in star clusters
leads to shorter lifetimes, and faster expansion, and has a stronger
impact on initially segregated clusters than on unsegregated
clusters. (Vesperini et al. 2009b).

Measured mass functions of a sample of globular clusters
surprisingly show that all high concentration clusters have steep
mass functions (i.e., larger $\alpha$), while low concentration ones
have a smaller $\alpha$ (De Marchi et al. 2007). This is puzzling,
because concentrated clusters are expected to be dynamically more
evolved and thus most of their low mass stars should have evaporated
from the clusters. However, Marks et al. (2008) showed that for
initially mass-segregated clusters, mostly low-mass stars are lost
due to gas expulsion which implies a shallower slope in the low-mass
range. Indeed this approach allows to deduce the evolution of the
very early Milky Way in unprecedented time resolution as it allows
us to deduce the changes in the cluster potential induced by
gas-expulsion and the varying Galactic tidal field \citep{Marks10}.

Several studies have been done to dynamically model the evolution of
star clusters using $N$-body simulations. Recently, Harfst et al.
(2009) modelled the very young Arches cluster to find the best
fitting initial model by comparing simulations with observational
data. In this way they aimed at constraining the parameters for the
initial conditions of the cluster such as IMF, size, and mass of the
cluster by neglecting the Galactic potential and stellar evolution.
Moreover, using $N$-body simulations, a close correlation between
the slope of the low-mass end of the stellar mass function and the
fraction of the initial cluster mass loss has been found by several
authors (Vesperini \& Heggie 1997; Baumgardt \& Makino 2003; Trenti
et al. 2010).

In the present paper we report on simulations of the diffuse outer
halo globular cluster Palomar 14 (Pal 14). Pal 14 with a mass of
about $10^4$ M$_{\odot}$ and a (three dimensional) half-light radius
of 34 pc, is a particularly interesting object to study. It is one
of the most extended Galactic globular clusters and, with a distance
of $\simeq 70$ kpc (Jordi et al. 2009) is also among the outermost
clusters of the Milky Way. This large Galactocentric distance and
its low stellar mass density make this cluster an ideal candidate to
decide between Newtonian gravity and MOND (Baumgardt et al. 2005,
Haghi et al. 2009). Its low density suggests that binary stars may
be an important issue for interpreting its measured velocity
dispersion \citep{Kuepper10}. Its low mass together with its large
radius make it also possible to simulate Pal~14 on a star-by-star
basis. However, simulations of the evolution of individual globular
cluster's long-term evolution have so far been done only with Monte
Carlo or Fokker Planck simulations \citep{gie09}. We therefore
performed a set of direct $N$-body simulations to determine the most
likely starting conditions of Pal~14 in terms of total mass, initial
half-mass radius and stellar mass function. We also test if
primordial mass segregation is necessary to reproduce the observed
flattened mass function. Moreover, we check if primordial binaries
get frequently disrupted within such a sparse cluster or whether
Pal~14 has preserved its primordial binary content. It should be
noted that all simulations performed in the present study start once
the cluster is gas-free and has settled back into virial
equilibrium.

This paper is organized as follows: in Section 2 we describe the
observational data we use to compare our simulations to. Our
methodology is described in Section 3. In Section 4 we explain the
different N-body set-ups and compare the numerical results with the
observed data. Conclusions are given in Section 5.

\section{Observational data}

We use observational data of Hilker (2006) and Jordi et al. (2009)
who have presented a spectroscopic and photometric study of Pal 14.
Jordi et al. (2009) have obtained the cluster's mass function using
HST/Wide Field Planetary Camera 2 (HST/WFPC2) data from the HST
archive. Moreover, they determined Pal 14's velocity dispersion by
measuring the radial velocities of 17 giant stars using the
Ultraviolet-Visual Echelle Spectograph (UVES; Dekker et al. 2000) at
the Very Large Telescope (VLT) of the European Southern Observatory
(ESO) in Chile and the High Resolution Echelle Spectrograph (HIRES;
Vogt et al. 1994) on the Keck I telescope.

The distance of Pal 14 from the Sun was derived from the apparent
magnitude of the horizontal branch in the colour-magnitude diagram
and an isochrone fit to be about $71\pm1.3$ kpc (Jordi et al. 2009).
This is closer to the Sun than previous estimates, e.g. Hilker
(2006) and Dotter et al. (2008) derived larger distances of 74.7 kpc
and 79 kpc, respectively. Pal 14 has a projected half-light radius
of $~1'.28"$ (Hilker 2006), which was derived from a curve-of-growth
analysis of the integrated light of the member stars. This
corresponds to a projected half-light radius of $R_{phl}=26.4\pm0.5$
pc at the assumed distance of $71\pm1.3$ kpc and implies a 3D
half-light radius of about  $35.6\pm 0.6$ pc.

Jordi et al. derived the age of the cluster to be 11.5$\pm$ 0.5 Gyr
by adopting [Fe/H] = -1.50 for the  metallicity (Harris 1996) from
the best isochrone fitting of Dotter et al. (2007). Although
previous work has shown that Pal 14 is 3-4 Gyr younger than typical
halo GCs with the same metallicity, thus has an age of $\simeq 10$
Gyr (Sarajedini 1997; Hilker 2006; Dotter et al. 2008), the recent
estimated age by Jordi et al. reduces this difference.

The density profile of the cluster does not exhibit a significant
ellipticity and the distribution of its stars is compatible with a
homogeneous, circular distribution within the uncertainties (Hilker
2006). The core radius, and the central surface brightness of the
cluster were determined from the best fitting King (King 1966)
profiles. The concentration parameter which is defined as $c = log
(r_t/r_c)$ is about 0.85 (Hilker 2006), i.e. rather low compared to
most Milky Way globular clusters.

The velocity dispersion of Pal 14 by measuring radial velocities of
17 red giant cluster members has been presented with and without
including one probably binary or AGB star (hereafter \texttt{star
15}), whose velocity is more than 3$\sigma$ away from the mean of
the other stars (Jordi et al. 2009). For the sample including
\texttt{star 15}, the global line-of-sight velocity dispersion is
0.64 $\pm$ 0.15 kms$^{-1}$, and without \texttt{star 15} it is 0.38
$\pm$ 0.12 kms$^{-1}$ (Jordi et al. 2009). The quoted uncertainties
reflect the 1-$\sigma$ error.

The upper boundary of Pal 14's main sequence corresponds to a
magnitude of 23.4 mag at a wavelength of 555 nm. Based on
photometric data, Jordi et al. (2009) used the masses given by the
11.5 Gyr isochrone by Dotter et al. (2007), and found that the
observed stellar masses on the MS covered the mass range of 0.495
$\mbox{M}_{\odot}$ up to 0.825 $\mbox{M}_{\odot}$. The mass function
of Pal 14 was found to be flatter than a Kroupa or Salpeter IMF in
the observed mass range, and the best fitting power-law mass
function was obtained for a slope of $\alpha =1.27 \pm 0.44$.

%===============================================================

\begin{figure}
\includegraphics[width=87 mm]{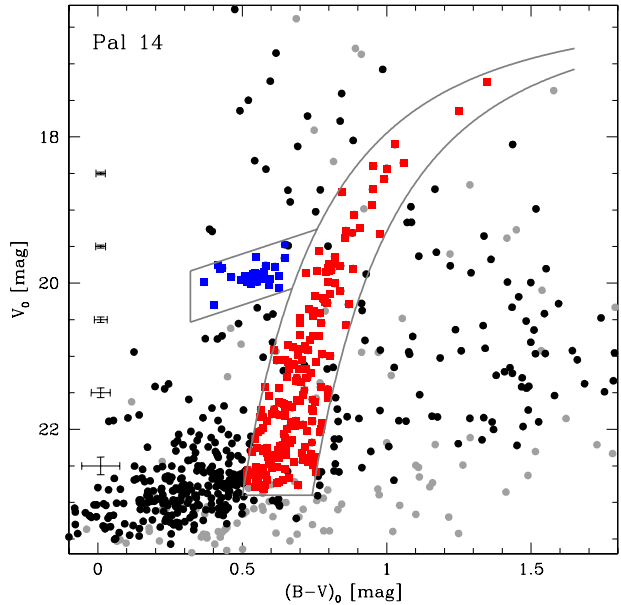}
\caption{Colour-magnitude diagram of Pal 14. The grey dots are all
photometric data, the black dots are a selection of point sources
with reasonable photometric errors ($<0.2$ mag). The lines give the
selection regions of RGB and HB stars. We put the bottom limit of
the RGB to $V_0 = 22.9$ mag, which would correspond to $M_{V,0} =
3.48$. Red squares are those stars in the RGB region that are within
the tidal radius of Pal 14. The same applies for the blue squares
for the HB stars. } \label{cmd}
\end{figure}

\begin{figure}
\includegraphics[width=87 mm]{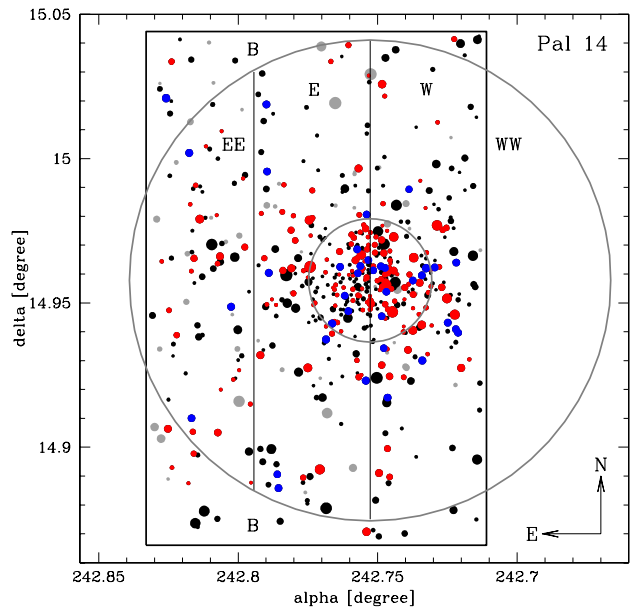}
\caption{ Spatial distribution of the photometrically observed Pal
14 stars. The big rectangle marks the area of the observations. The
red dots are all objects in the RGB regime shown in the CMD, the
blue dots are the objects in the HB area of the CMD. The two large
circles are the half-light radius at 1.15 arcmin and the tidal
radius of 4.81 arcmin (both from Baumgardt et al. (2005)). The
region is divided into stripes/segments, called EE, E, W and WW from
East to West. Everything outside the tidal radius is the background
(B) which was used to count the contaminating objects. } \label{coo}
\end{figure}

The total mass of stars inside the half-light radius and within the
mass range 0.525 $\mbox{M}_{\odot}$ to 0.825 $\mbox{M}_{\odot}$ has
been estimated to be $M= (2200\pm 90) \mbox{M}_{\odot}$. With the
measured slope for masses down to 0.5 $\mbox{M}_{\odot}$ and using a
Kroupa-like mass function, $\alpha=1.3$ for masses between 0.1
$\mbox{M}_{\odot}$ and 0.5 $\mbox{M}_{\odot}$, Jordi et al. (2009)
estimated the total mass of Pal 14 to be about $M_{tot}=12000
\mbox{M}_{\odot}$ without taking into account stellar remnants.
Since this estimated total mass is a derived quantity and not
directly measured, we use the number of bright main-sequence stars,
$N_{bs}$ to compare with our models, instead. For this aim, we used
the star-count data with completeness factor greater than 0.50,
which leads to the mass range from 0.525 M${\odot}$ to 0.795
M${\odot}$ and consequently to the number of Pal 14's bright stars
$N_{bs}=2954$ (Table 3 of Jordi et al. (2009)).  This mass range
corresponds to a magnitude range $27.2> m_{555}> 23.5$. $m_{555}$ is
the HST filter, F555W, that is very similar to Johnson V (Fig. 7 of
Jordi et al. (2009)). The stellar mass at the main-sequence turn-off
point, $m_{MSTO,0}=23.44$, is about $0.79$ M$_{\odot}$
\citep{dot07}. Consequently the central and average density of the
cluster are $0.1$ and $0.01$ stars/$pc^3$, respectively. The
two-body relaxation time of Pal~14 is about 50 Gyr, i.e. several
times larger than a Hubble time.

The total number of giant stars including the red giant branch (RGB)
and horizontal branch (HB) stars is also another observable
quantity. The photometry published by Hilker (2006) has not covered
entirely the cluster (a segment in the West was missing) and was not
corrected for foreground/background contamination. Therefore, the
numbers in the tables of Hilker (2006) are not complete. We
therefore redo a new selection and incompleteness calculation to
count all RGB and HB stars. Figure \ref{cmd} shows Pal 14's
colour-magnitude diagram (CMD). Note that some of the brighter RGB
stars on the bluer side might actually be AGB stars. This cannot be
differentiated from this photometry. From this CMD it also becomes
clear that it makes no sense to count SGB stars. The photometric
error is large at those magnitudes and the counts are highly
incomplete. The spatial distribution of the observed stars is shown
in Figure \ref{coo}. We divided the regions into stripes/segments,
called EE, E, W and WW from East to West. Everything outside the
tidal radius is the background (B) which was used to count the
contaminating objects. Obviously, the missing region WW should
contain more or less the same number of RGB/HB stars as the region
EE. But one can also make other sanity checks of number counts, for
example whether there are the same amount of RGB/HB stars in the E
and W stripes, or whether really the half-light radius divides the
number of RGB/HB stars into half (assuming no mass segregation).

The total area within the tidal radius ($r<r_t$) is $78.5$
arcmin$^2$. There are different ways to derive the number of RGB and
HB stars: 1) assuming that there are as many RGB stars in the WW
segment as in the EE segment, 2) doubling the number of RGB stars in
the East (E+EE). The first method gives N$_{RGB}= 164\pm18$, the
second N$_{RGB}= 169\pm24$, agreeing very well with the first
number. Both values give the background subtracted number counts,
taking into account the ratio of the selected area vs. the
background area, i.e. scaling the background counts to the selected
area. This analysis shows that one can safely assume that the RGB
counts in the West should be similar to those in the East. Another
result is that there are slightly more RGB stars inside the
half-light radius, N$_{RGB}= 46\pm7$, than outside the half-light
radius, N$_{RGB}= 39\pm10$. The numbers are consistent within the
errors, but might hint to mass segregation of RGB stars on the East
side. By the same analysis for the HB stars in the tidal radius we
find  N$_{HB}= 34 \pm7$ for method 1) and N$_{HB}= 33 \pm10$ for
method 2), both values being in very well agreement. The total
number of giant stars (RGB+HB), N$_g = 198 \pm 19$, is given in the
tables to compare with those obtained from the modelled clusters.

%===============================================================
In order to reproduce those observational values, e.g. the
half-light radius, $R_{phl}$, the number of bright stars, $N_{bs}$,
and the mass function slope, $\alpha$, we construct a set of
$N$-body models for Pal 14 in the next section.

\section{$N$-body Models}
In total we computed a set of 66 models to find the initial
conditions which reproduce the observations of Pal 14 best. The
simulations were carried out with the collisional $N$-body code
\textsc{NBODY6} (Aarseth 2003) on the GPU computers of the Stellar
Population and Dynamics Group of the University of Bonn.

All clusters were set up using the publicly available code
\textsc{McLuster}\footnote{\tt
www.astro.uni-bonn.de/\~{}akuepper/mcluster/mcluster.html}(K\"upper
et al. (2010), in preparation). We used a Plummer profile in virial
equilibrium as initial mass distribution and computed all clusters
for 11 Gyr. We chose the initial mass and size (i.e. half-mass
radius) of the cluster as free parameters and kept the other
parameters of the initial conditions fixed. The number of stars was
varied in the range $7\times10^4<$ N $<10^5$ such that the clusters
have total masses in the range of $4.6-5.4\times10^4$ M$_{\odot}$.
Since the half-mass radius grows due to stellar-evolutionary mass
loss and dynamical evolution over the cluster life time, we chose
the initial 3D half-mass radii in the range of 18 pc to 22 pc to
reach projected half-light radii of about 26 pc after 11 Gyr.

We have chosen two different initial mass functions for our models.
First, we start with the canonical IMF (Kroupa 2001) using lower and
upper mass limits of 0.08 M$_{\odot}$ and 100 M$_{\odot}$,
respectively. At the second step, the clusters were constructed
using a flatter mass function with lower and upper mass limits of
0.08 M$_{\odot}$ and 5 M$_{\odot}$, respectively, mimicking the
initial conditions of an initially mass-segregated star cluster
after gas expulsion at an age of about 100 Myr. Stellar evolution
was modelled according to the routines by Hurley et al. (2000). Due
to the low escape velocity of Pal 14, we assume a 0\% retention
fraction for neutron stars and black holes which form during the
simulation. Note that the canonical (universal) two-part power-law
IMF has $\alpha_1=1.3$ for $0.08\leq m \leq 0.5$ M$_{\odot}$ and
$\alpha_2=2.3$ for $m\geq 0.5$ M$_{\odot}$. Here we refer to
$\alpha$ as the index of the mass function in the stellar mass range
0.525 M${\odot}$ to 0.795 M${\odot}$.

For simplicity, and since no information on the proper motion is
available, the clusters move on a circular orbit through a
logarithmic potential, $\phi(R_{G})= V_{G}^{2}$ ln$R_{G}$, of the
Milky Way at the assumed distance of Pal 14 and with a circular
velocity $V_{G}$ = 220 km/sec.

All but one cluster do not have primordial binaries. We constructed
one model with a primordial binary fraction of 5\% to study the
impact of binaries on cluster evolution and to estimate the fraction
of binaries which dissolve in a Hubble time in a cluster as sparse
as Pal 14. The binary population was set up using the Kroupa (1995)
period distribution with random pairing of the binary companions.
Random pairing of companion masses for primaries less massive than 1
M$_{\odot}$ is the correct procedure to reproduce the pre-main
sequence and main sequence binary-star data (Kroupa 2008).

The modelled clusters were assumed to be not initially mass
segregated. In addition we computed a number of clusters with
different degrees of primordial mass segregation to investigate the
effect on the cluster's evolution. An overview of the clusters is
given in Tables 1, 2 and 3.

\section{Finding the best fitting model}\label{best-fit}

%=======================================================================================================  Fig 1
\begin{figure*}
\begin{center}
\includegraphics[width=168mm]{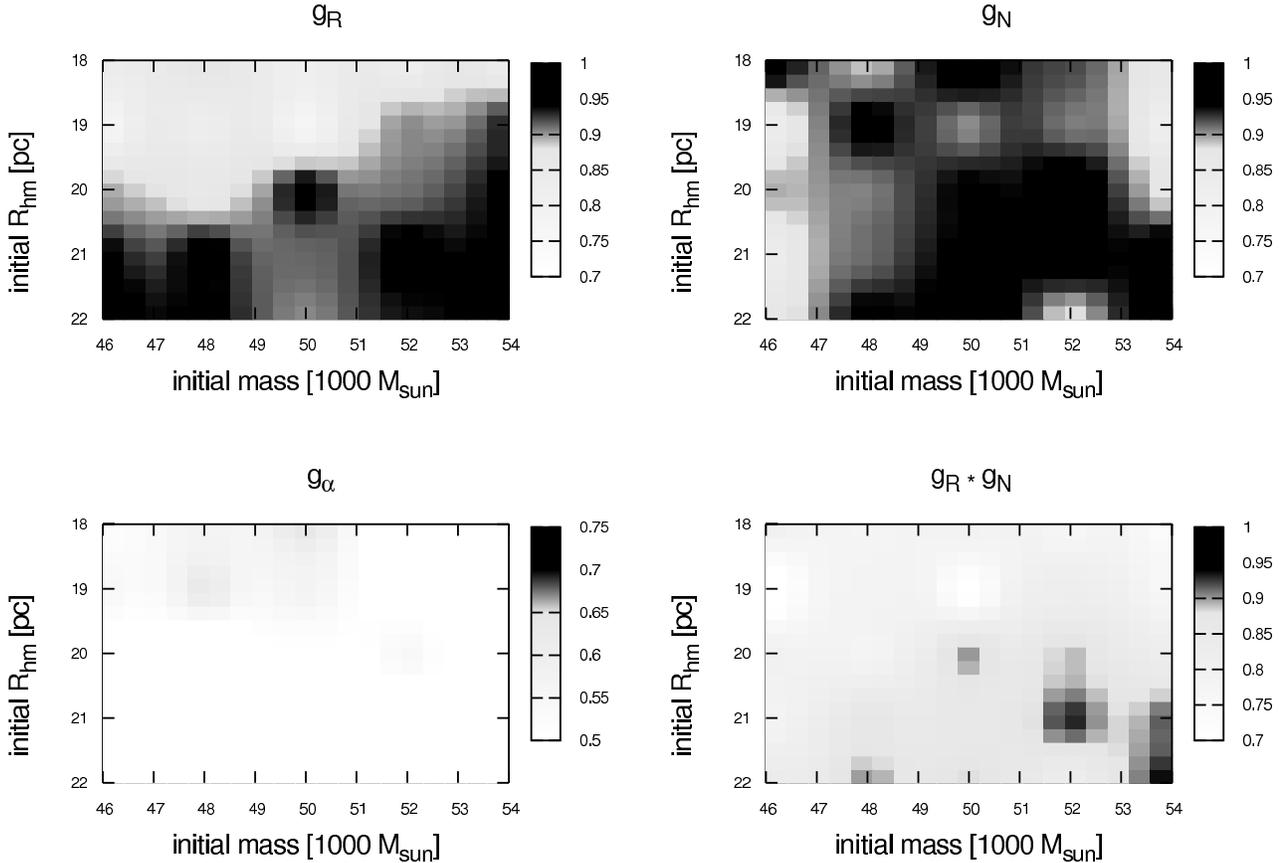}
\caption{Goodness-of-fit parameters from the measured half-light
radius (upper left panel), number of bright stars (upper right
panel) and mass function slope (lower left panel) for the models
with regular initial conditions. In the lower right panel the
product of the two upper panels is given. The varying model
parameters are the initial half-mass radius and the cluster initial
mass. A goodness-of-fit parameter close or equal to 1 is best and
indicated by dark shaded areas in the plots.} \label{gof_regular}
\end{center}
\end{figure*}
%=============================================================================================================
In order to find the most likely initial conditions, a grid of
models with different initial half-mass radii and initial masses was
calculated. The main observational values which were used to compare
with each model, are the present day number of bright main-sequence
stars, $N_{bs}$, the present day projected half-light radius,
$R_{phl}$, and the slope of the mass function in the mass range
0.525 M${\odot}$ to 0.795 M${\odot}$ within the half-light radius,
$\alpha$, with values
\begin{eqnarray}
N_{bs}&=&2954 \pm 175 ,\\
R_{phl}&=&(26.4 \pm 0.5)\, \mbox{pc}\\
\alpha &=&  1.27 \pm 0.44,
\end{eqnarray}

Note that, to match $R_{phl}$ in the $N$-body models we use only the
giant stars since the projected half-light radius comes from Hilker
(2006) and he could only see the giant stars in Pal 14. Moreover,
for $N_{bs}$ we count the main-sequence stars inside the half-light
radius with masses between 0.525 $\mbox{M}_{\odot}$ and 0.795
$\mbox{M}_{\odot}$. Finally, we extract the mass function for those
main sequence stars within the half-light radius, with a mass in the
range of $0.525 \mbox{M}_{\odot}< m < 0.795 \mbox{M}_{\odot}$, i.e.
in the same way as Jordi et al. have done. We apply 10 bins between
$\log_{10} m = -0.1$ and $\log_{10} m = -0.3$. We reject the
lowermost point in the observational data of Jordi et al. since it
has a relatively high incompleteness correction, i.e. uncertainty.
Since compact remnants are unobservable in Pal 14, as they are too
faint, we ignored them in our measurement of the mass function.

To compare our models in an objective way and to measure the quality
of the fit of the model to the observations, we define a
goodness-of-fit parameter, $g_{p}$. This parameter depends on the
model parameters, and measures the closeness between the
observations and the model as follow
\begin{equation}
 g_{p}=1-\left|\frac{p_{observation}-p_{model}}{p_{observation}}\right|, \label{fitness}
\end{equation}
where $p$ represents the corresponding model parameter. The
goodness-of-fit parameter ranges from 0 to 1, where 1 describes an
excellent fit. Since we have three main observables, $R_{phl}$,
$N_{bs}$ and $\alpha$, we also have three goodness-of-fit
parameters, $g_{R}$, $g_N$ and $g_\alpha$, respectively.

There is a significant statistical dispersion in the results if we
repeat the calculations for one arbitrary model with the same
initial parameters but with a different random seed. The required
computational GPU time doesn't allow us to do statistics for all
parameters, though. Hence, in order to estimate the errors of the
resulting parameters, we do several runs (10 times) for a particular
model and discuss the resulting uncertainty. In the following
investigation we will use these uncertainties for our discussion.

In the following sections we will investigate four sets of models.
First, we investigate the initial conditions which we call
'regular'. We assume a Kroupa IMF without primordial mass
segregation or primordial binaries. In the second step we assume
that the IMF was flattened within the first 100 Myr through stellar
evolution and gas expulsion processes. The third set of models is
computed to test how much primordial mass segregation we have to add
in order to explain the observed mass function of Pal 14. In the
last section we test the influence of primordial binaries on the
cluster evolution and on the evolution of the mass function.

\subsection{Regular initial conditions}\label{sec:regular}

%================================================================================================================== Fig 2

\begin{figure}
\includegraphics[width=87 mm]{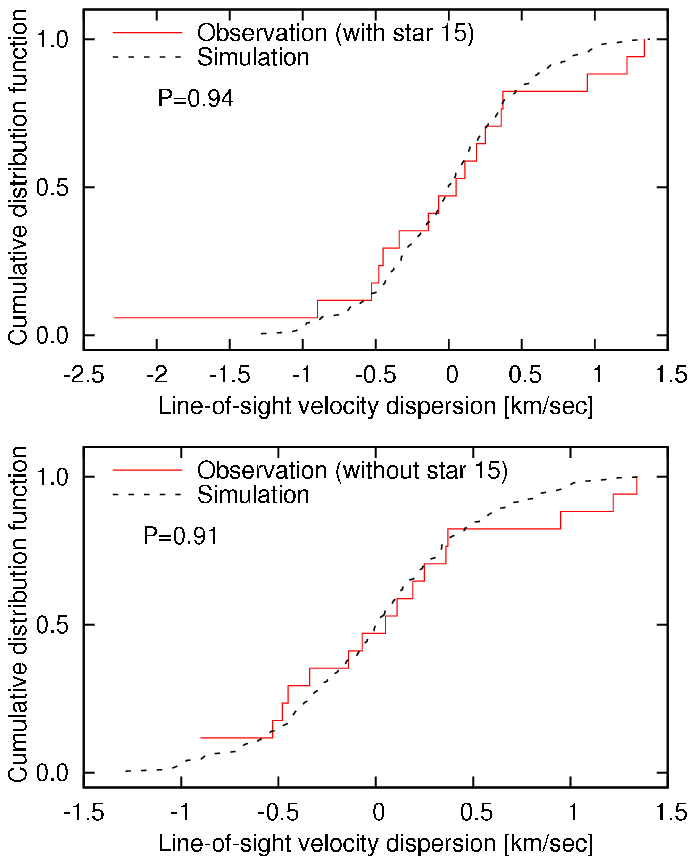}
\caption{ The cumulative distribution function (cdf) versus
line-of-sight velocity for the sample with \texttt{star 15} (upper
panel) and without \texttt{star 15} (bottom panel). The solid lines
represent the observed data, whereas the dotted line is the cdf of
one of the modelled clusters (M50R20). The maximum difference
between the two cdfs is 0.13 for the upper panel and 0.14 for the
bottom panel, which corresponds to $P$-values of 0.94 and 0.91,
respectively.  } \label{ks}
\end{figure}

%===============================================================================================================  Fig 3

\begin{figure}
\includegraphics[width=87mm]{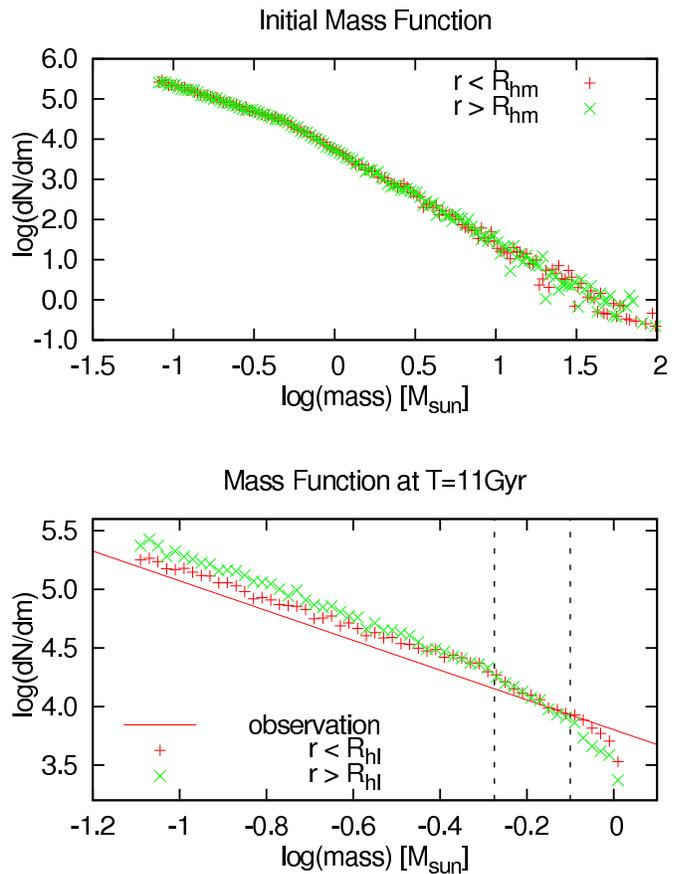}
\caption{Mass function of main-sequence stars for one of the
modelled clusters, 'M50R20', inside (red pluses) and outside (green
crosses) the half-light radius, $R_{hl}$. The top panel shows the
unsegregated mass function at the start of the simulation which was
chosen to be the canonical Kroupa IMF. The observed mass function
slope inside the half-light radius is indicated by the solid line.
The dotted vertical lines indicate the observed range of stellar
mass. The bottom panel shows the mass function after an evolution of
11 Gyr. The slope of the mass function in the observed range is
$1.97\pm0.15$ which is larger than the observed value. The flattened
slope of the low-mass part and the steeper slope of the high-mass
part within the half-light radius indicate that mass segregation has
happened in the cluster.} \label{mf_regular}
\end{figure}

%========================================================================================================    Tab 1
\begin{table*}
\begin{minipage}{144mm}
\centering \caption{Details of all $N$-body models with regular
initial conditions (Sec.~\ref{sec:regular}). The first column gives
the model name, in which the first two digit numbers denote the
initial mass in 1000 M$_{\odot}$ and the second two digit numbers
denote the initial half-mass radius in parsec. Columns 2 and 3 are
the projected half-mass and half-light radius, $R_{phm}$ and
$R_{phl}$, respectively. Column 4 is the present day number of
bright main-sequence stars inside the half-light radius, with masses
between 0.525 and 0.795$\mbox{M}_{\odot}$, $N_{bs}$, whereas column
5 gives the corresponding total mass of these stars. $\alpha$ is the
present day slope of the mass function for stars with masses between
0.525 and 0.795$\mbox{M}_{\odot}$. Column 7 is the line-of-sight
velocity dispersion. The last column gives the total number of giant
stars in the cluster including RGB and HB stars. Typical errors of
the numerical values were obtained by repeating runs for a
particular model and are indicated in the header. In the last line,
the observational values are given for comparison. The subscript
'$*$' indicates the observed velocity dispersion with taking
\texttt{star 15} into account. Note that none of these models are an
acceptable fit to Pal 14. }
\begin{tabular}{cccccccc}
\hline
Model &$R_{phm}$[pc]&$R_{phl}$[pc]&$N_{bs}$&$M^{f}_{r<R_{hm}}$[M$_{\odot}$]&$\alpha$&$\sigma_{los}$[km/sec]&$N_{g}$\\
&($\pm$0.8)&($\pm$2.3)&($\pm$260)&($\pm$320)&($\pm$0.15)&($\pm$0.01)&($\pm$16)\\
\hline M40R15     &19.6&15.6&2374&1521&2.10&0.65&162\\
       M40R25     &31.5&30.3&2182&1402&1.85&0.51&136\\
       M46R18     &23.5&22.0&2912&1869&1.96&0.64& 163\\
       M46R19     &24.3&19.8&2382&1530&1.90&0.64&161\\
       M46R20     &25.8&23.8&2661&1705&2.04&0.61&174\\
       M46R21     &28.4&25.2&2407&1542&2.08&0.58&188\\
       M46R22     &29.3&26.8&2544&1627&2.10&0.58&160\\
       M48R18     &23.7&23.1&3311&2127&1.88&0.66&174\\
       M48R19     &24.3&21.5&2847&1821&1.77&0.64&160\\
       M48R20     &26.2&22.5&2641&1689&2.13&0.61&172\\
       M48R21     &28.0&25.6&2687&1722&2.02&0.59&174\\
       M48R22     &28.1&27.6&3079&1970&2.14&0.61&171\\
       M50R18     &24.2&21.3&2992&1925&1.78&0.68&194\\
       M50R19     &24.5&19.5&2638&1695&1.87&0.65&193\\
       M50R20     &26.6&25.8&3108&1995&1.97&0.64&170\\
       M50R21     &26.9&24.1&2796&1788&2.14&0.61&178\\
       M50R22     &29.3&29.0&2997&1918&2.10&0.60&163\\
       M52R18     &24.3&21.5&3181&2038&2.06&0.68&181\\
       M52R19     &25.7&24.0&3245&2079&2.00&0.66&186\\
       M52R20     &26.6&24.0&2975&1910&1.90&0.65&166\\
       M52R21     &28.8&27.0&3027&1939&2.05&0.62&172\\
       M52R22     &28.0&29.0&3357&2155&1.98&0.61&171\\
       M54R18     &24.0&21.8&3369&2158&2.07&0.70&224\\
       M54R19     &25.5&24.8&3556&2279&2.00&0.66&195\\
       M54R20     &26.3&25.1&3348&2144&2.08&0.65&161\\
       M54R21     &27.9&25.1&3017&1935&1.96&0.63&190\\
       M54R22     &30.0&26.0&2865&1834&2.15&0.62&198\\
       M60R25     &32.1&33.0&3699&2367&2.22&0.62&170\\
    \hline
    Observations    &  &26.4$\pm$0.5& 2954$\pm$175  &2200$\pm$90& 1.27$\pm$0.44 & 0.38$\pm$0.12 & 198$\pm$19    \\
       &    &    &    &   &   &(0.64$\pm$0.15$^{\ast}$)&   \\
       \hline
\end{tabular}
\label{tab_regular}
\end{minipage}
\end{table*}

Figure \ref{gof_regular} shows how the goodness-of-fit parameters of
$N_{bs}$, $R_{phl}$, and $\alpha$ vary for the different values of
the model parameters, initial half-light radius and initial cluster
mass. Moreover, the product of the goodness-of-fit parameters for
$N_{bs}$ and $R_{phl}$ is shown in the lower right panel.

The darker areas show where the goodness-of-fit parameters are close
to 1 in the parameter space, i.e. show the range in initial
conditions that best fit to Pal 14. We find that many models (e.g.
M54R22, M52R21, and M50R20) can reproduce Pal 14's half-light radius
and its number of bright main-sequence stars.

We look for models whose velocity dispersion lies within the
uncertainties of the observational values. Therefore, we take only
stars within a clusters tidal radius and with a mass higher than 0.8
M$_{\odot}$, and measure their velocity dispersion and compare it to
the observed value of \cite{jor09}. According to Table
\ref{tab_regular}, the line-of-sight velocity dispersion in most of
our models is about $0.6$ km s$^{-1}$.

Given the small sample size (i.e., radial velocity of 17 stars are
used to determine the observed velocity dispersion), the observed
velocity dispersion has a large error which could be much larger
than the formal error (Gentile et al. 2010). In order to see how
well a velocity dispersion of  $\simeq0.6$ km s$^{-1}$ agrees with
the observed one, we use the Kolmogorov-Smirnov (KS) test, which is
the applicable statistical test for such a small sample size.

In Fig. \ref{ks} we show the cumulative distribution function (cdf)
vs. radial velocity for both samples with and without \texttt{star
15}, separately. This cdf we compare to the cdf of one of the best
fitting models (M50R20). With \texttt{star 15}, the maximum
difference between the two cdfs is 0.13, which corresponds to
P-value of 0.94, while in case without star 15 the maximum
difference between the two cdfs and corresponding P-value are 0.14
and 0.91, respectively. This means that the models can be rejected
with 6\% (including star 15) and 9\% (excluding star 15) confidence.

In addition, we find that the mean value of the velocity dispersion
of all stars is similar to that of the red giant stars and main
sequence stars with masses larger than 0.8 M$_{\odot}$. Therefore,
the velocities of the giants are representative of all stars in the
cluster. The number of giant stars, $N_g$, is also in most of the
modelled clusters in good agreement with the observations.

Figure \ref{mf_regular} shows the mass function of one example model
before and after the computation. The red and green points denote
the number of stars inside and outside the half-light radius,
respectively. We see that after 11 Gyr the mass function inside the
half-light radius is significantly different from the one outside
the half-light radius, which implies that mass segregation has taken
place. The mass function of the stars inside the half-light radius
is clearly flatter than those of the stars outside. It seems that
even in a cluster as large as Pal 14, dynamical mass segregation
plays an important role.

According to Table \ref{tab_regular} and lower left panel of Figure
\ref{gof_regular}, the slope of the mass function, $\alpha$, in all
models is far from the observed value of Pal 14\textbf, though, even
when accounting for observational errors. The upper limit of the
observed value within the 1-$\sigma$ error is $\alpha=1.71$, and
only if we account for the statistical error for the modelled slope,
which is $\simeq$0.15, then two models (i.e., M48R19 and M50R18) are
marginally compatible with the observed slope.

Our simulations have typically $\alpha\simeq2.0$ after 11 Gyr , i.e.
only a mild decrease of the low-mass-star slope from the initial
Salpeter value of $\alpha=2.35$ due to mass segregation. In
contrast, the observed slope is $\alpha=1.27 \pm 0.44$, which agrees
with this value only at the 1.75 $\sigma$ level. There is therefore
some evidence that Pal 14 started with a mass segregated Kroupa IMF.

In order to see whether an extension of the initial parameter space
would resolve the problem, we repeat the simulation for some
extremely high and low initial values, i.e. we chose 25 pc and 15 pc
for the half-light radius and 40000 M$_\odot$ and 60000 M$_\odot$
for the initial mass. The results are shown in Table
\ref{tab_regular}. As can be seen, even by selecting such values for
the initial conditions, the present day mass function of Pal 14
cannot be reproduced. Moreover, we find that the values for $N_{bs}$
and $R_{phl}$ get much worse in these directions of the parameter
space, showing that we were searching in the right place initially.

\subsection{Flattened mass function}\label{sec:flat}

%===============================================================================================================
\begin{table*}
\begin{minipage}{159mm}
\centering \caption{ Details of all $N$-body models for three sets
of flattened mass functions, with initial slopes of 0.5, 0.6, and
0.7 for stars less massive than 0.5 M$_\odot$ (see Sec.
\ref{sec:flat} for more details). Columns are the same as in Table
\ref{tab_regular} but in column 2, the first number in the model
name is the adopted high-mass slope. Errors are the standard
deviations derived from multiple runs with the same parameters (see
text). The best fitting models that agree within the uncertainties
with all observational parameters are highlighted with boldface.}
\begin{tabular}{ccccccccc}
\hline
&Model &$R_{phm}$ [pc]&$R_{phl}$ [pc]&$N_{bs}$&$M^{f}_{r<R_{hm}}$[M$_{\odot}$]&$\alpha$&$\sigma_{los}$[km/sec]& $N_{g}$\\
&&($\pm$1.6)&($\pm$2.3)&($\pm$290)&($\pm$270)&($\pm$0.13)&($\pm$0.03)&($\pm$25)\\
\hline
         1&F0.5M45R18     &26.7 & 24.9 & 2391 & 1543 & 1.19 & 0.59  & 216\\
         2&F0.5M45R20     &28.7 & 27.5 & 2322 & 1476 & 1.32 & 0.56  & 200\\
         3&F0.5M45R22     &30.8 & 29.5 & 2332 & 1551 & 1.53 & 0.53  & 216\\
         4&\bf{F0.5M50R18}     &26.9 & 27.7 & 3045 & 1959 & 1.28 & 0.62  & 253\\
         5&\bf{F0.5M50R20}     &29.1 & 27.7 & 2635 & 1678 & 1.31 & 0.60  & 234\\
         6&F0.5M50R22     &30.9 & 30.6 & 2722 & 1718 & 1.16 &  0.57 & 242\\
         7&\bf{F0.5M55R18}     &26.0 & 26.3 & 3327 & 2120 & 1.38 &  0.66 & 285\\
         8&\bf{F0.5M55R20}     &29.4 & 27.2 & 2824 & 1819 & 1.22 &  0.62 & 259\\
         9&F0.5M55R22     &30.5 & 29.3 & 2926 & 1917 & 1.38 &  0.60 & 268\\
         \hline
         10&\bf{F0.6M45R18}     &25.2 & 25.0 & 2878 & 1903 & 1.25&  0.61 & 218\\
         11&F0.6M45R20     &27.9 & 25.5 & 2493 & 2017 & 1.36&  0.59 & 228\\
         12&F0.6M45R22     &30.0 & 28.5 & 2506 & 1646 & 1.60&  0.56 & 210\\
         13&\textbf{F0.6M50R18}     &25.9 & 23.9 & 2987 & 1999 & 1.35&  0.64 & 249\\
         14&\bf{F0.6M50R20}     &27.7 & 28.0 & 3133 & 2001 & 1.47&  0.61  & 261\\
         15&F0.6M50R22     &31.1 & 31.5 & 3132 & 2162 & 1.51&  0.60  & 217\\
         16&F0.6M55R18     &25.4 & 23.5 & 3248 & 2189 & 1.26&  0.68  & 264\\
         17&\bf{F0.6M55R20}     &27.7 & 26.1 & 3358 & 2155 & 1.38&  0.64  & 296\\
         18&F0.6M55R22     &30.5 & 29.6 & 3257 & 2002 & 1.36&  0.63  & 275\\
         \hline
         19&F0.7M45R18     &24.3 & 21.0 & 2734 &  1759 &  1.43&  0.68 & 238\\
         20&\bf{F0.7M45R20}     &26.6 & 25.6 & 2971 &  1926 &  1.47&  0.61 & 232\\
         21&F0.7M45R22     &29.2 & 29.6 & 3021 &  1937 &  1.74&  0.59 & 222\\
         22&F0.7M50R18     &24.4 & 22.1 & 3204 &  2101 &  1.25&  0.68 & 279\\
         23&\bf{F0.7M50R20}     &26.5 & 25.1 & 3324 &  2172 &  1.77&  0.65 & 283\\
         24&F0.7M50R22     &28.8 & 28.0 & 3419 &  2202 &  1.47&  0.69 & 272\\
         25&F0.7M55R18     &24.4 & 21.5 & 3516 &  2209 &  1.40&  0.72 & 330\\
         26&F0.7M55R20     &26.6 & 25.5 & 3773 &  2526 &  1.41&  0.68 & 307\\
         27&F0.7M55R22     &29.6 & 27.5 & 3451 &  2290 &  1.45&  0.66 & 289\\
      \hline
         &Observations    &  &26.4$\pm$0.5& 2954$\pm$175&2200$\pm$90& 1.27$\pm$0.44 &0.38$\pm$0.12  & 198$\pm$19     \\
         &    &    &     &           & &          &(0.64$\pm$0.15$^{\ast}$)&\\
         \hline
\end{tabular}
\label{tab_flat}
\end{minipage}
\end{table*}

%============================================================================================================
\begin{figure}
\includegraphics[width=87mm]{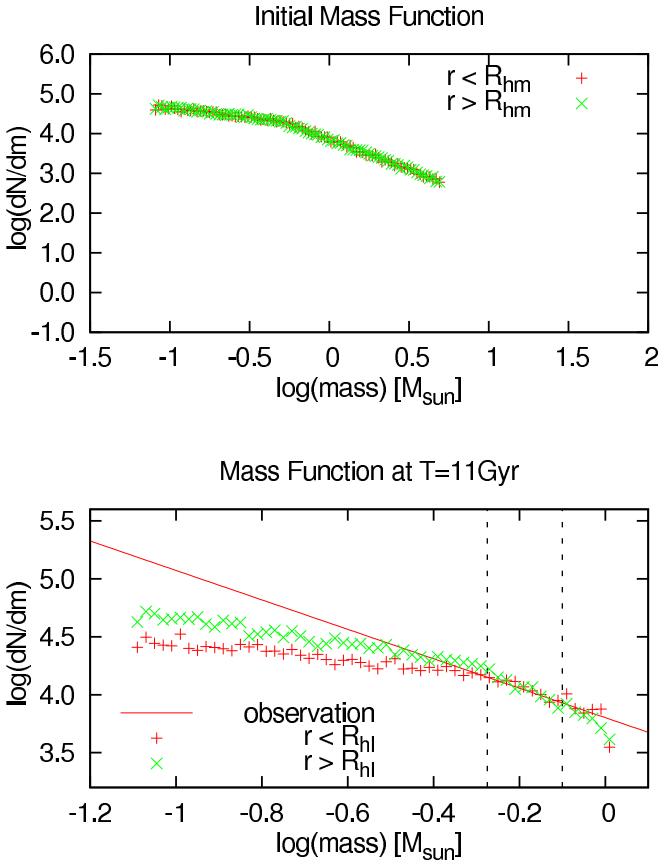}
\caption{ The same as Figure \ref{mf_regular}, but here we started
with a flattened mass function with a slope $\alpha = 1.5$ above
0.5M$_{\odot}$, and $\alpha = 0.5$ below 0.5M$_{\odot}$. The initial
mass of this particular model is 50000 M$_{\odot}$ and the initial
half-light radius is 18 pc. The maximum mass in the mass spectrum
was set to 5 M$_{\odot}$, instead of 100 M$_{\odot}$. Further
properties of the cluster after 11 Gyr of evolution are given in
Table \ref{tab_flat}. } \label{mf_flat}
\end{figure}

%============================================================================================================
\begin{figure}
\includegraphics[width=93 mm]{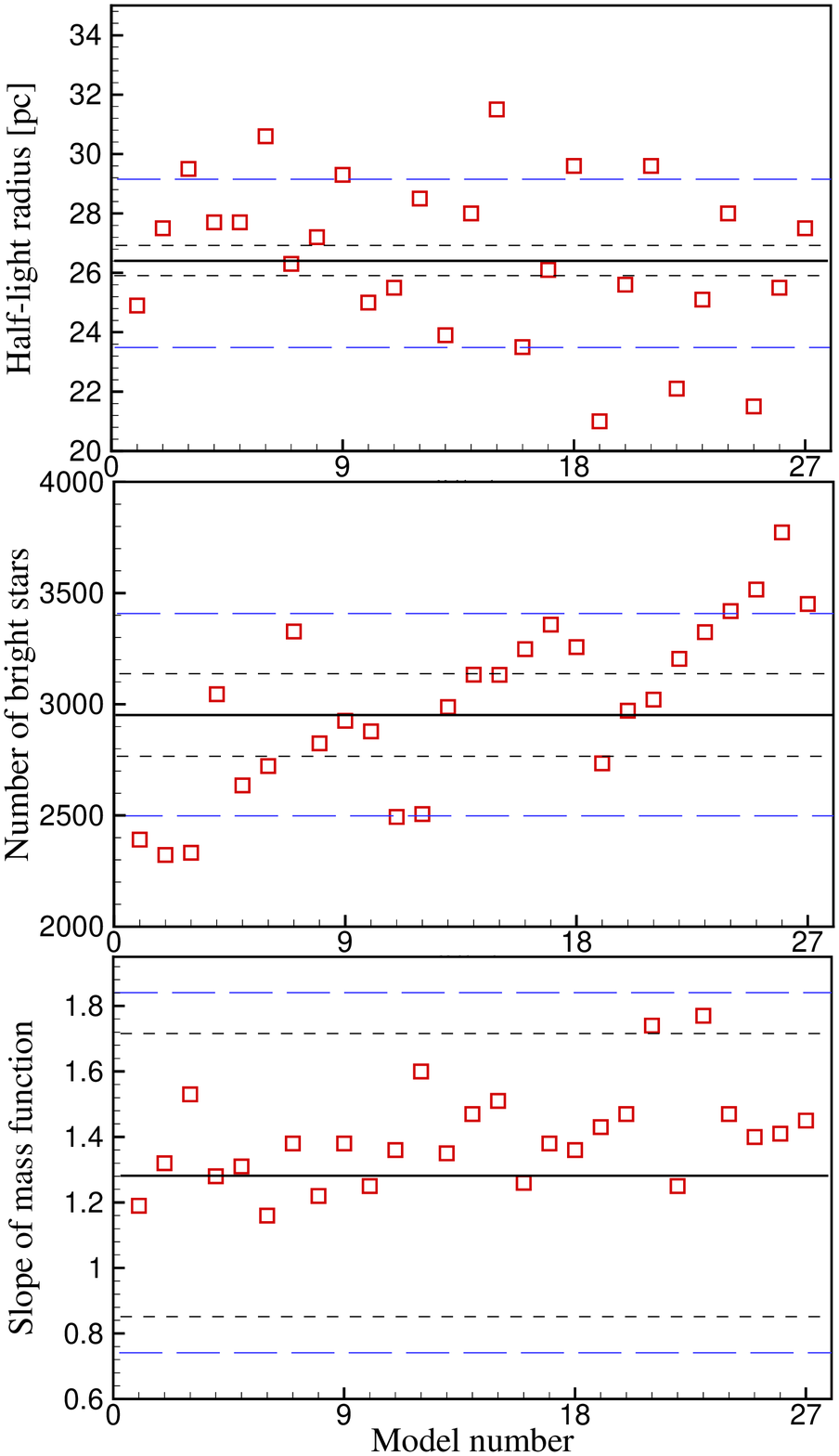}
\caption{  The half-light radius (top panel), the number of bright
stars (middle panel), and the present day slope of the mass function
for stars with masses between 0.525 and 0.795$\mbox{M}_{\odot}$
(lower panel) for all models with flattened IMFs (Sec.
\ref{sec:flat}). The horizontal axis is the model number listed in
Table \ref{tab_flat}. The horizontal solid and dashed lines indicate
the observed values and the corresponding errors, respectively. The
horizontal long-dashed lines show the total errors including the
typical numerical errors given in Table \ref{tab_flat}, i.e. points
outside these lines are not compatible with the observations.
Further properties of the clusters after 11 Gyr of evolution are
given in Table \ref{tab_flat}. } \label{flat}
\end{figure}

%=======================================================================================================

As shown in Section \ref{sec:regular}, the initially unsegregated
clusters with a canonical Kroupa IMF cannot reproduce the observed
slope of the mass function well. The observed slope is
$\alpha=1.27\pm0.44$ but the slope that could be achieved from
$N$-body calculations is about 2. In this section we show that if we
choose a flatter IMF for our model, it is possible to explain the
observed slope of the mass function.

The evolution and disruption of star clusters due to gas expulsion
and stellar evolution may play an important role in the evolution of
the globular cluster mass function (GCMF). It has been shown that
these early evolutionary processes, i.e. early cluster dissolution
due to gas expulsion and mass loss from stellar evolution, can
preferentially destroy low-mass clusters and significantly flatten
the low-mass end of the power-law initial GCMF ( Goodwin 1997,
Goodwin 1998, Kroupa \& Boily 2002, Baumgardt et al. 2008b,
Parmentier et al. 2008, Vesperini 2009c). That is, the birth process
of a globular cluster can be very violent, depending on the exact
initial conditions of its parent gas cloud \citep{Baumgardt07,
Baumgardt08b, par07}. Therefore, the first few million years can
have a significant impact on a globular cluster's evolution over a
Hubble time. \cite{Marks08} find that, depending on the
concentration of the initial cluster, gas expulsion can induce a
flattening of the stellar mass function of the cluster.
Unfortunately, this process cannot be easily computed in a direct
way for a globular cluster of the pre-gas expulsive mass of Pal 14,
because clusters can lose a large fraction of their birth stellar
population as a result of gas expulsion and stellar evolution mass
loss. We therefore have to use an approximation here.

Thus, following \cite{Marks08} and assuming that a certain
flattening of the mass function slope has happened within the first
100 Myr of the cluster's evolution, we change the initial mass
function slope and start the simulations at a cluster age of 100
Myr. In this fashion we have performed a new series of $N$-body
simulations for models with various flattened initial slopes of the
mass function instead of the canonical Kroupa IMF (see overview in
Table \ref{tab_flat}). We have chosen three sets of slopes for the
mass function: $\alpha_a=\{1.7,0.7\}$, $\alpha_b=\{1.6,0.6\}$, and
$\alpha_c=\{1.5,0.5\}$, where the first number in each set is the
slope of the mass function for stars more massive than 0.5
M$_{\odot}$ and the second one is for low-mass stars (for
comparison, the Kroupa IMF is $\alpha = \{2.3,1.3\}$). The second
column in Table \ref{tab_flat} shows the name of the simulated
models. For example, F0.6M50R20 represents a flattened model with
the slope of $\alpha_b$, initial mass of 50000 M$_{\odot}$, and
initial half-mass radius of 20 pc.

The maximum mass in the mass spectrum was set to 5 M$_{\odot}$,
instead of 100 M$_{\odot}$. The reason can be attributed to the
short lifetimes of the most massive stars, i.e., the stars with
masses larger than 5 M$_{\odot}$ will have died or turned into
compact remnants within the first 100 Myrs. As stated earlier, it is
reasonable to assume a retention fraction of 0\% for compact
remnants due to the low escape velocity from Pal 14.

In Figure \ref{mf_flat} we plot the mass function evolution for one
of the best fitting models, F0.5M50R18. We see that mass segregation
has taken place, in addition to our artificial flattening of the
mass function, by about the same amount as in the non-flattened case
in Fig.~\ref{mf_regular}.

The results are summarized in Table \ref{tab_flat}. We find that a
number of models in Table \ref{tab_flat} can reproduce the present
day properties of Pal 14. From Fig. \ref{flat} we see that all
models with a flattened IMF can reproduce the observed slope of the
mass function much better. However, the half-light radius, due to
the small value of the observational error, can constrain the
calculated models. The number of bright stars is also well
reproduced in some models. As can be seen in Fig. \ref{flat}, models
with a lower mass and slope pair $\alpha_a$, as well as models with
a higher mass and slope pair $\alpha_c$ are not compatible with the
observed value of $N_{bs}$.

In order to calculate the uncertainties of our results we have
simulated each model for several times (i.e., 2 to 6 runs for each
model). This also gives us some insight to the uncertainties of the
models in Section \ref{sec:regular}. The resulting errors are also
given in Table 2. These are about the same as the errors of the
models in Table 1.

Like in Section \ref{sec:regular}, the line-of-sight velocity
dispersion in most of our models is about $0.6$ kms$^{-1}$ and is
compatible with the observed velocity dispersion including or
excluding \texttt{star 15}.

As can be seen in Table \ref{tab_flat}, due to the larger number of
massive stars in the clusters with initially flattened
mass-function, the  average values of $N_g$ are larger than for
clusters with regular initial condition. However, the number of
giant stars in some of the best fitting models (e.g. F0.5M50R20,
F0.6M45R18, F0.6M50R18, and F0.7M45R20) are still compatible with
the observed value within the uncertainties.

\subsection{Primordial mass segregation}\label{sec:seg}

%============================================================================================================
\begin{figure}
\includegraphics[width=87 mm]{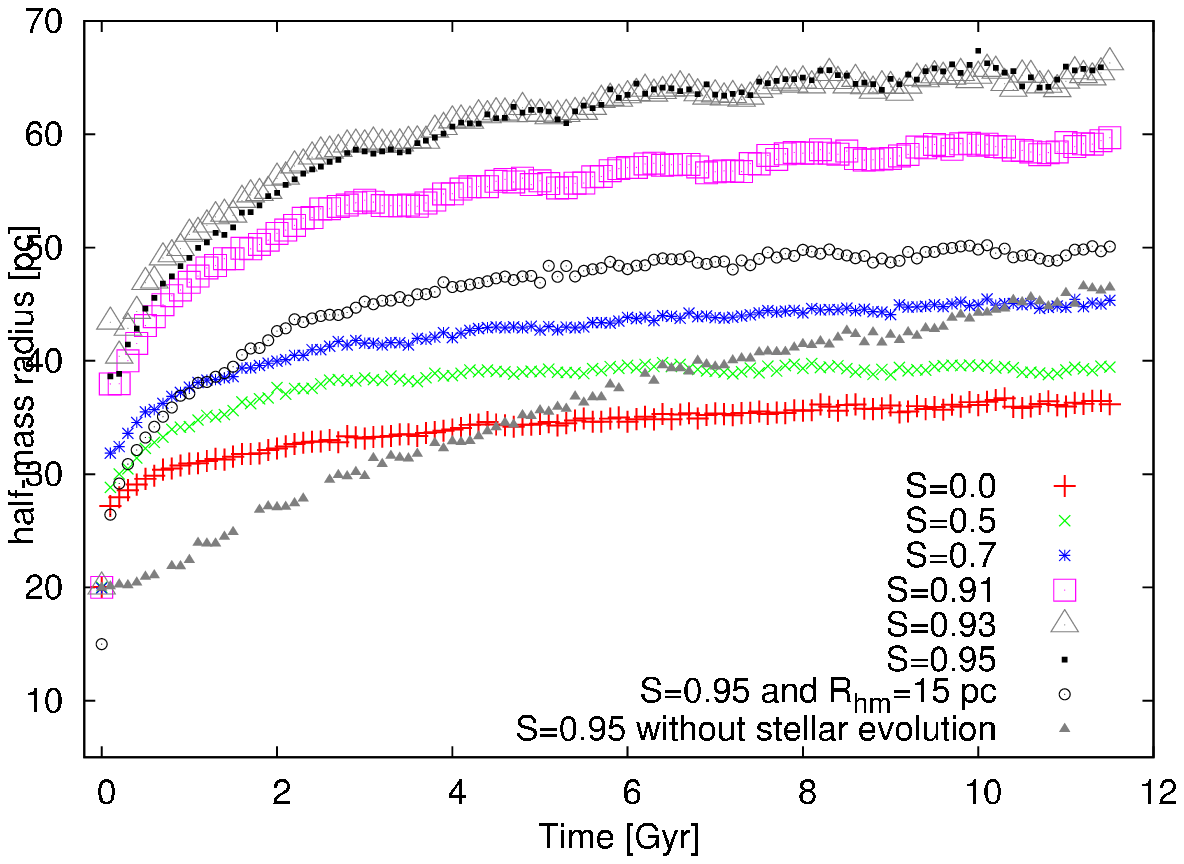}
\caption{ Evolution of the 3D half-mass radius for different degrees
of primordial mass segregation with a canonical Kroupa IMF. The
initial mass is 50000 M$_{\odot}$, and the initial half-mass radius
is 20 pc for all models, except for one model with $R_{hm}=15 pc$.
Due to stellar evolution clusters experience a rapid expansion such
that clusters with higher degrees of primordial mass segregation
experience a larger jump in the half-mass radius within the first
100 Myr of evolution. For comparison, the stellar evolution was
artificially switched off for one model. As can be seen, this
cluster has no sudden expansion in the beginning of its evolution.}
\label{rh_evol}
\end{figure}
%=============================================================================================================

\begin{table*}
\begin{minipage}{151mm}
\centering \caption{ Details of all $N$-body models with primordial
mass segregation and of the one model without primordial mass
segregation but with primordial binaries. The mass segregation
parameter changes in the range $S=0.5-0.95$; the binary fraction for
the binary model is 0.05. Columns are the same as in Table
\ref{tab_regular}, just in column 1 the first number within the name
of the model is here the adopted mass segregation parameter, $S$, or
the fraction of stars in binaries, respectively. The best fitting
models that agree within the uncertainties with all observational
parameters are highlighted with boldface.}
\begin{tabular}{cccccccc}
\hline
Model               &$R_{phm}$ [pc]&$R_{phl}$ [pc]&$N_{bs}$&$M^{f}_{r<R_{hm}}$[M$_{\odot}$]&$\alpha$&$\sigma_{los}$[km/sec]&$N_{g}$\\
&$\pm$1.2&$\pm$1.9&$\pm$120&$\pm$80&$\pm$0.10&$\pm$0.05&$\pm$8\\
\hline S0.50M50R20  &29.4&25.3&3184&2044&1.90&0.62&193\\
       S0.60M50R20  &32.2&27.5&3209&2061&1.9&0.58&207\\
       S0.70M50R20  &32.9&27.9&3267&2106&2.0&0.60&191\\
       S0.80M50R20  &35.6&29.6&3219&2066&1.95&0.57&222\\
       S0.90M50R20  &42.4&36.0&3257&2100&1.6&0.51&211\\
       S0.91M50R20  &43.8&38.5&3365&2169&1.63&0.46&206 \\
       S0.93M50R20  &47.8&40.6&2887&1870&1.27&0.44&217\\
       S0.95M50R20  &48.7&41.5&3159&2045&1.28&0.47&215\\
       \textbf{S0.95M50R15}  &37.0&27.6&3077&1978&1.33&0.58& 194\\
       \textbf{S0.90M50R15}  &33.9&26.3&3236&2085&1.67&0.55& 210\\

    \hline
        B0.04M50R20 & 27.4 & 25.3 & 3132 & 2009 &2.0& 0.87& 185\\
    \hline
    Observations    &  &26.4$\pm$0.5& 2954$\pm$175  &2200$\pm$90& 1.27$\pm$0.44 &   0.38$\pm$0.12& 198$\pm$19    \\
       &    &    &    &   &   &(0.64$\pm$0.15$^{\ast}$)&   \\
       \hline
\end{tabular}
\label{tab_seg}
\end{minipage}
\end{table*}

%===================================================================================================
\begin{figure}
\includegraphics[width=87mm]{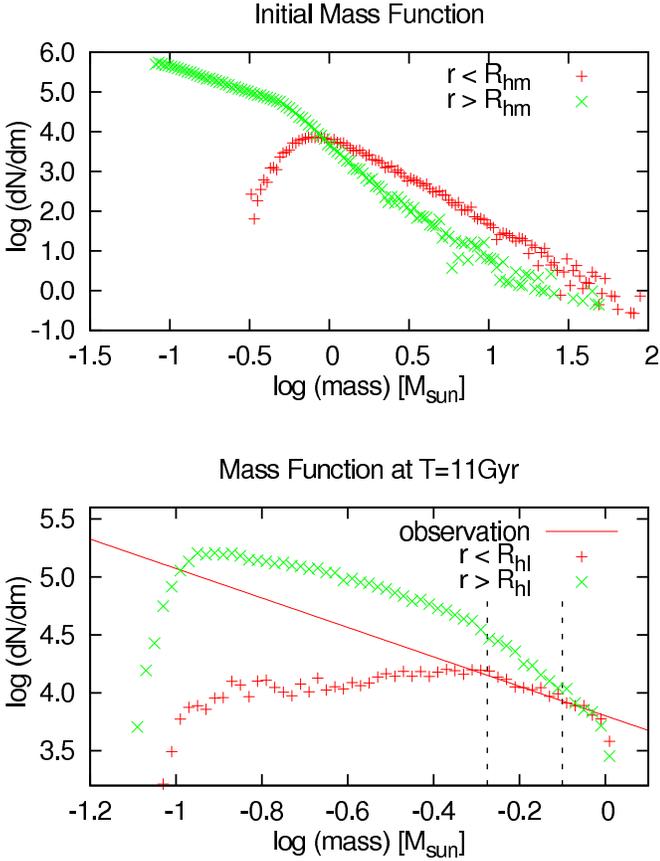}
\caption{ The same as Figure \ref{mf_regular}, but here we started
with a primordially mass segregated cluster with a canonical Kroupa
IMF. The initial mass of this particular model (S0.95M50R15) is
50000 M$_{\odot}$, the initial half-light radius is 15 pc, and the
mass segregation parameter is set to $S=0.95$. Clusters with such a
strong degree of primordial mass segregation are able to reproduce
the observed flat mass function inside the half-light radius (solid
line).
 } \label{mf_seg}
\end{figure}

%===================================================================================================================
\begin{figure}
\includegraphics[width=87 mm]{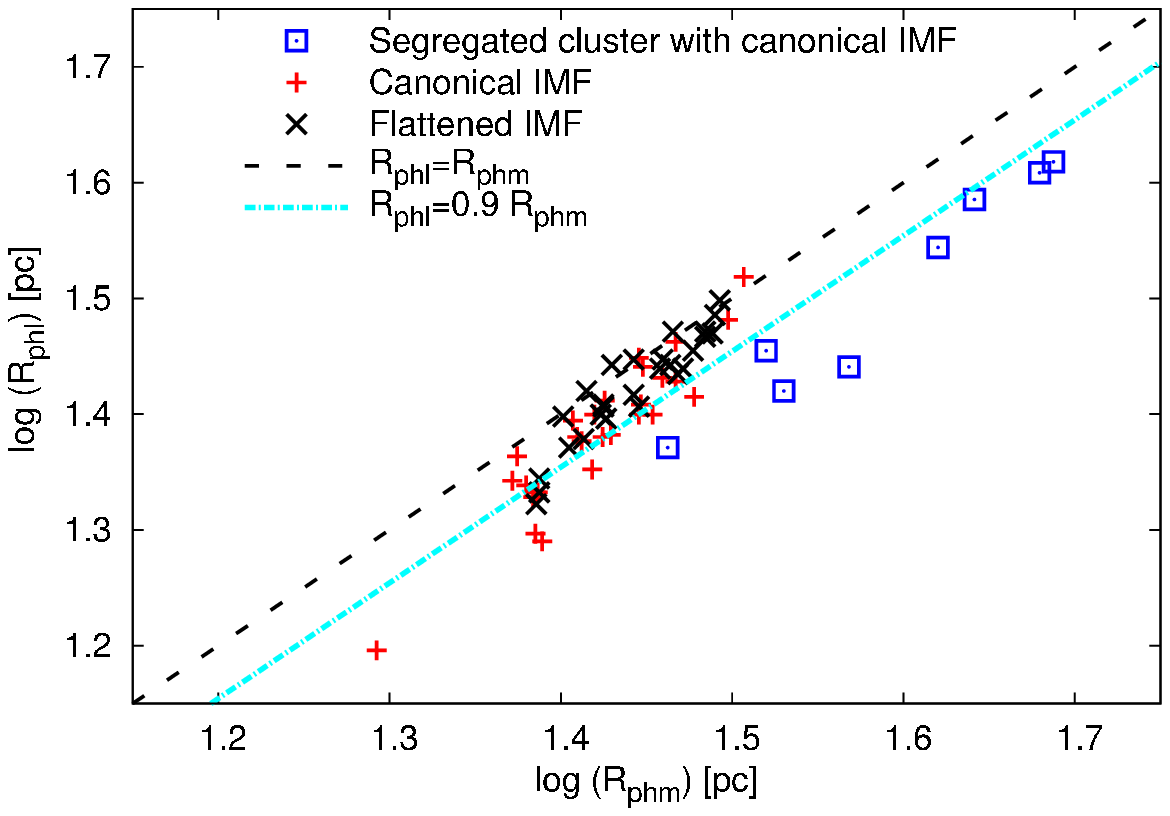}
\caption{ The projected half-light radius ($R_{phl}$) plotted
against the projected half-mass radius ($R_{phm}$) for the different
sets of models listed in Tables 1-3. The black dotted line shows the
region where $R_{phl}=R_{phm}$, whereas the cyan dash-dotted line
demonstrates if $R_{phl}$ is 10\% smaller than $R_{phm}$. Initially
segregated clusters show more than 10\% difference between $R_{phl}$
and $R_{phm}$. } \label{Rml}
\end{figure}
%==========================================================================================

Another possibility to explain the fact that a standard IMF gives a
poor fit to the observed mass function of Pal 14 is the assumption
that Pal~14 was primordially mass segregated. To test this
hypothesis, we set up clusters with various degrees of mass
segregation. The code \textsc{McLuster} allows to specify a degree
of mass segregation parameter (hereafter, $S$). $S=0$ means no
segregation and $S=1$ refers to full segregation, where the most
massive star is in the lowest point of the cluster potential and the
least massive star is in the highest point of the cluster potential
\citep{sub08}. \textsc{McLuster} uses the routine described in
\citet{Baumgardt08a} to segregate the clusters. This routine allows
to maintain the desired density profile when increasing the degree
of mass segregation while also asserting the cluster to be in virial
equilibrium.

For this part of the investigation we take one of our modelled
clusters of Sec. \ref{sec:regular} and add different degrees of
primordial mass segregation. Thus, the initial 3D half-mass radius
and initial cluster mass are fixed to $M=50000 \mbox{M}_{\odot}$ and
$R_{hm}=20$ pc, respectively. The mass segregation parameter was
changed in the range $S=0.5-0.95$.

The results of the simulated models are shown in Table
\ref{tab_seg}. In all cases the mass function flattens as the
cluster loses stars. The amount of mass function flattening depends
on the given amount of mass segregation, i.e. initially more
segregated clusters have more flattened mass functions than
unsegregated clusters. From Table \ref{tab_seg} we see that low
values of $S$ cannot reproduce the observed mass function slope,
while highly segregated clusters ($S\geq0.9$) are able to fit the
observed slope of the mass function.

We also find that primordial mass segregation in clusters leads to a
stronger expansion than for unsegregated clusters. Figure
\ref{rh_evol} shows the evolution of 3D half-mass radius of clusters
for different values of the mass segregation parameter. We see that
by increasing primordial mass segregation, the final half-mass
radius rises. For example, for a mass segregation degree varying
between 0.5 to 0.95 and an initial half-mass radius $R_{hm}=20$ pc,
the final 3D half-mass radius changes from 39 pc to 67 pc. This
expansion is due to both dynamical and stellar evolution. The
stellar evolution in the first 100 Myr has a more important effect
and leads to a jump in the cluster's radius such that the cluster
with a higher degree of segregation experiences a larger jump.
Consequently, this increase of radius leads to a lower value of the
mean velocity dispersion.

In order to reproduce the final $R_{phl}$ compatible with the
observed value, we therefore have to choose a value for the initial
3D half-mass radius as small as $R_{hm}=15$ pc for a mass
segregation parameter of $S\geq0.9$. In this case, the final
line-of-sight velocity dispersion in this model is about 0.6
kms$^-1$, which is compatible with the observed value. Figure
\ref{mf_seg} shows the evolution of the mass function of this
particular model. The mass function slope after 11 Gyr of evolution
reproduces the observations reasonably well. This model can also
reproduce all other observational parameters well, but such a high
value for the mass segregation parameter implies a very tight
correlation of a star's mass and its specific energy within the
cluster at birth. There is no conclusive evidence for such a high
degree of primordial mass segregation in any observed star cluster.
In addition, as can be seen in Fig. \ref{mf_seg}, the difference
between the slope of the mass function inside and outside the
half-light radius is larger than for other models. Thus, further
analysis to determine the observed mass function outside the
half-light radius would help to distinguish the mass segregation
scenario from other scenarios.

Figure \ref{Rml} compares the projected half-light radii, $R_{phl}$,
and the projected half-mass radii, $R_{phm}$ for the three different
sets of modelled clusters from Sec. \ref{sec:regular}-\ref{sec:seg}.
As can be seen, in almost all models $R_{phl}$ is smaller than
$R_{phm}$. For initially segregated clusters this difference is more
than 10\% such that the assumption that mass traces light within a
cluster is disputable. It must be mentioned that 10\% difference in
cluster radius leads to 30\% difference in the number of stars,
which is significant and should be taken into account in
discussions.

\subsection{Primordial binaries}\label{sec:bin}
%======================================================================================================== Fig 9
\begin{figure}
\includegraphics[width=87 mm]{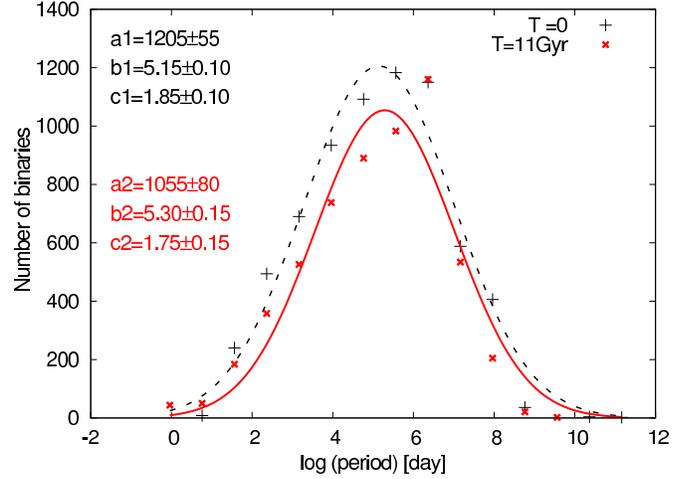}
\caption{ The period distribution of bound binary systems found in
the cluster at the beginning and after 11 Gyr evolution. Dashed and
solid lines are the best Gaussian fits to the initial and final
period distributions, respectively. The coefficients of the Gaussian
fits with a functional form of $a\exp[(x-b)^2/2c^2]$ are given in
the figure. The errors give the 1 $\sigma$ errors. The period
distribution stays the same within the errors, showing that
dynamical encounters are unimportant in changing the binary
population in Pal 14. Almost 85\% of primordial binaries have
survived the cluster evolution.} \label{period}
\end{figure}

Observations and theoretical work suggest that stars may be born
with initially very large binary fractions and thus star clusters
may contain a significant population of primordial binaries
\citep{kro95a,kro95b}. Low density stellar systems (i.e., less than
$10^2$ M$_{\odot}$pc$^{-3}$) have a binary fraction of at least
50\%, while in dense clusters (i.e., more than $10^4-10^5$
M$_{\odot}$pc$^{-3}$), the binary fraction is less than 10-20\%
\citep{hut92}.

The stellar density in Pal~14 is extremely low (0.1-0.2
M$_{\odot}$pc$^{-3}$), and thus it is expected that most of the
primordial binaries survive during the evolution of the cluster.
Thus, the present day binary population could be close to the
primordial binary population which it had at birth
\citep{kro95a,kro95b,Kuepper10} making Pal 14 a good target to
determine the primordial binary fraction of a globular cluster. Here
we discuss one model with primordial binaries.

Since tight binaries can significantly speed up the relaxation
process, one might be able to explain the flattening of the observed
mass function, since by adding binaries, one increases the mean mass
of particles in the cluster (as a binary acts like one heavy
particle), and by increasing the mean mass we decrease the
relaxation time \citep{heg03}.

Unfortunately, binaries slow down direct $N$-body computations
immensely because time steps have to be very small for their
integration, such that most numerical investigations neglect
binaries completely. It is only the peculiar configuration of Pal 14
with its relatively low mass and low density that enables us to
perform such a simulation with a few thousand binaries over a
time-span of 11 Gyr. Yet, such a computations takes about one month
GPU time on a GeForce GTX 280 GPU.

We set up one of the modelled clusters of Sec. \ref{sec:regular}
with an initial mass of M=50000 M$_{\odot}$, initial half-mass
radius of $R_{hm}=20$pc, a canonical Kroupa IMF without primordial
mass segregation but with a non-zero primordial binary fraction. We
add a few binaries to see if the present day binary distribution is
a good match for the primordial one. A primordial binary fraction,
\begin{equation}
f_{bin}=\frac{N_{bin}}{(N_{bin}+N_s)}, \label{fbin}
\end{equation}
of 0.05 (i.e. about 10\% of stars being in binaries because $N_s\gg
N_{bin}$), where $N_{bin}$ is the number of binary systems in the
cluster and $N_{s}$ is the number of single stars, is adopted, in
order to see how many of them will be destroyed over the course of
time and which effect on the cluster's evolution they have. The
initial period distribution (i.e. the post-gas-expulsion period
distribution) used in this study is the birth period distribution of
Kroupa (1995b, 2008). Note that the birth period distribution is not
a Gauss function. But the period distribution function of bound
binaries (i.e., the widest binary systems are excluded) that results
from this birth distribution is approximately a Gauss distribution
already at $T=0\,$Gyr because the widest initial binaries are split
due to crowding even in the extremely low density of our initial Pal
14 model.

Our simulation shows that such a small binary fraction does not
increase the mass segregation enough, as 5\% binaries do not
increase the mean mass significantly. Hence, the obtained slope of
the mass function, which is about $\alpha\simeq2.0$ is still not
compatible with the observed value, $\alpha=1.27\pm0.44$ within a
1-$\sigma$ level of confidence.

After 11 Gyr evolution of the cluster, the binary fraction reaches
$f_{bin}\simeq0.04$, which means that during the evolution most of
the binaries ($\simeq$85\%) survived. In other word, the present day
binary fraction that we see in Pal 14 is almost equal to the
primordial binary fraction at its birth. Figure \ref{period} shows
the period distribution of binary systems at the beginning and after
11 Gyr of evolution. The parameters of the best-fitting Gaussian
distributions to the initial and final period distributions are
presented in Fig. \ref{period}. Since the mean period does not
change with time (within the error), most of the missing binaries
might have escaped the cluster rather than gotten disrupted.

Moreover, adding the binaries increases the mean velocity dispersion
such that it reaches $0.87$ km/sec if we consider all giant stars
and main sequence stars with mass higher than $0.8$ M$_{\odot}$. We
measure $0.64$ km/sec, if we add a cut-off in velocities leaving out
the stars that are $5$ km/sec away from the mean and hence would not
be considered as cluster members in a cluster like Pal 14. It should
be noted that the $5$ km/sec limit is only applied for this test to
compare the velocity dispersion of our model clusters with the
observations since observers would not count stars with such
discrepant velocities as members. In all other cases stars were
considered as members only based on their distance to the cluster
center. An overview of the results for this cluster can be found in
Table \ref{tab_seg}. In order to see how well these velocity
dispersions are compatible with the observed sample including
\texttt{star 15}, we use a KS test again. Here, the $P$-value is
$0.68$ for both cases with and without adding the cut-off for
high-velocity stars. This means that the data presented in Jordi et
al. (2009) is compatible with our modelled cluster with primordial
binaries with 68\% confidence. This is compatible with recent
results by \citet{Kuepper10}.

\section{Conclusions}

We performed a series of $N$-body computations in order to create a
realistic model for the outer halo globular cluster Pal 14 and show
that it is possible nowadays to directly calculate the evolution of
small Milky Way globular clusters over a Hubble time by direct
N-body simulations. For the sake of comparison, and as a rough
estimate, the modelling of a typical halo globular cluster (i.e.
with the initial mass of about $2\times10^5$ M$_{\odot}$ and
half-light radius of about 5 pc) over 12 Gyr will take about $10^5$
times longer than Pal 14, given the same resources. Moreover, about
half of the Galactic halo globular clusters are located at $R_G < 6$
kpc, hence the tidal field and cluster evaporation would be
stronger, thereby implying a higher initial cluster mass and longer
time of simulation.
%and show that it is possible nowadays to directly calculate the evolution of
%Milky Way globular clusters

Pal 14's large radius together with its low-mass, thus low-density
allow us to simulate it on a star-by-star basis. In this way we are
not only able to learn more about the dynamical history of such kind
of globular clusters, but are also able to study the process of
cluster formation as this period leaves its imprint on its ensuing
evolution. Moreover, we show that there is valuable synergy between
simulations and observations, as simulations can make predictions
for observations and, on the other hand, can test observational
parameters on their consistency.

Therefore, we generated 66 models divided into four categories:
clusters with a canonical Kroupa IMF (Sec.~\ref{sec:regular}),
clusters with a flattened IMF (Sec.~\ref{sec:flat}), clusters with a
Kroupa IMF but with primordial mass segregation
(Sec.~\ref{sec:seg}), and finally one cluster with primordial
binaries (Sec.~\ref{sec:bin}). By varying the initial half-mass
radius and the initial mass of the model clusters, we searched
through parameter space to obtain the model that matches the
available observations of Pal~14 best.

We used the most recent observational data to constrain the modelled
clusters. Thus, for all simulated clusters we compared the
half-light radius, the total number of bright stars inside the
half-light radius, and the present day mass function slope inside
the half-light radius in the mass range $0.5-0.8$M$_{\odot}$ with
the observed values. Furthermore, we checked the line-of-sight
velocity dispersion and the number of giant stars for their
consistency with the numerical results.

From the computations we find that all clusters segregate
significantly over the course of time, although the relaxation time
of Pal~14 is longer than a Hubble time. We furthermore find that
this mass segregation leads to an increasing flattening of the mass
function inside half-light radius, which is what we were looking for
since the observational value for the mass function slope of Pal~14
in the respective regime is only $\alpha = 1.27\pm0.44$
\citep{jor09} compared to a Kroupa IMF value of 2.3. The initially
unsegregated clusters with a canonical Kroupa IMF do not produce
enough depletion in the slope of the mass function in order to be in
good agreement with the observations, though. Only a few models of
this set marginally agree within the uncertainties with this value
for $\alpha$. Besides that, the other observational constraints are
matched reasonably well by most of the models in this set (Table
\ref{tab_regular} and Fig. \ref{gof_regular}).

The models with the flattened initial mass functions can make up for
the discrepancy in $\alpha$,  though some of them fail in the other
parameters (Table \ref{tab_flat} and Fig. \ref{flat}). Primordial
mass segregation also increases the flattening of the mass function
but the necessary degree of primordial mass segregation turns out to
be very high (Table \ref{sec:seg}).

The initially flattened mass function can be explained by loss of
low-mass stars from an initially mass-segregated cluster across the
tidal radius during the gas-expulsion phase \citep{Marks10}.
Therefore, the flattened initial mass function of Pal 14 we uncover
here strengthens the idea that this cluster formed in a stronger
tidal field rather than in isolation. This may have been the case if
either the Galactic tidal field has since then evolved, or if the
cluster was born closer to the Galactic centre (eccentric orbit),
or, more speculatively, the cluster birth site was a now
detached/disrupted dwarf galaxy. However, if Pal 14 has always been
on the same circular orbit, it is hard to imagine that the tidal
field would have played a role during the cluster response to gas
expulsion (Parmentier \& Kroupa 2010). Primordial binaries
($f_{bin}=0.05$) do not help to reproduce the observational value of
the mass function.

Thus, we are able to find some scenarios that can reproduce all
observations reasonably well. We find that the initial mass of Pal
14 was about 50000 M$_{\odot}$ and that it started with a half-mass
radius of about 20 pc. The cluster must have been significantly
mass-segregated with stellar mass function depleted in low-mass
stars. These are the signatures of a post-gas expulsion cluster
\citep{Marks08}. The inferred post-gas expulsion "initial" half-mass
radius is larger than the half-mass radii of most globular clusters
which is of the order of 3 pc. Moreover, we show that the different
scenarios show different mass functions outside the half-light
radius, such that finding the observational value for this region
would help to discriminate between the different scenarios.

The effect of unresolved binaries is an important issue that has to
be considered in the observational values of the velocity dispersion
and the mass function. If Pal 14 has a significant binary
population, which we expect due to the low density of this cluster,
the true (binary-corrected) mass function should be steeper, because
of the existence of unresolved components \citep{kro91}. Therefore,
the slope of the mass function is the most uncertain parameter of
the observations of Pal 14. Even without taking binaries into
account Jordi et al. (2009) estimate the uncertainty to be
$\simeq\pm$0.4. Furthermore, as K\"{u}pper \& Kroupa (2010) have
shown, if the binary fraction was normal (i.e. more than 50\%), the
binary-corrected velocity dispersion of the cluster would also be
smaller than the value reported by Jordi et al. (2009).

From the set of computations we furthermore find that primordial
mass segregation leads to a larger impact of stellar evolution on
cluster expansion. This is, segregated clusters are more affected by
mass loss through stellar evolution (Fig. \ref{rh_evol}). We
furthermore quantify the effect of mass segregation on the ratio of
half-mass radius to half-light radius and find the latter to be up
to 10\% smaller (Fig. \ref{Rml}), leading to significant
observational uncertainties for the number of stars within this
radius in such cases.

Moreover, we measured the line-of-sight velocity dispersions of our
models taking into account all stars, and compared it with the
observed values. We additionally measured the velocity dispersion
considering only giant stars, and find that both values do not show
significant difference between each other, showing that giant stars
are good tracers of a cluster's global velocity dispersion. Most
clusters show a velocity dispersion of $\simeq0.6$ km/sec,
suggesting that \texttt{star15} of the sample of \citet{jor09} might
be a normal cluster member.

Finally we find from the model with a primordial binary fraction of
5\% that, due to the low stellar density of Pal~14 the initial
binary distribution stays almost unchanged by the dynamical
evolution of the cluster, and only about 15\% of binaries escape or
get disrupted in the course of time (Fig. \ref{period}). Low-density
clusters like Pal 14 are therefore good objects to study the
primordial binary distribution of globular clusters.

\section*{Acknowledgements}
A.H.Z and H.H thank the Stellar Population and Dynamics Group of the
Argelander Institute for Astronomy and the Institute for Advanced
Studies in Basic Sciences (IASBS) for providing financial support
for this research.

\bsp \label{lastpage} 
\begin{thebibliography}{99}

\bibitem[\protect\citeauthoryear{Aarseth}{2003}]{ars03}
Aarseth S. J., 2003, Gravitational N-Body Simulations, Cambridge, UK, Cambridge University Press

\bibitem[\protect\citeauthoryear{Anders et al.}{2009}]{and09}
Anders, P.; Lamers, H. J. G. L. M.; \& Baumgardt, H. 2009  A\&A, 502, 817.

\bibitem[\protect\citeauthoryear{Baumgardt \& Makino}{2003}]{bau03}
Baumgardt H., Makino J., 2003, MNRAS, 340, 227.

\bibitem[\protect\citeauthoryear{Baumgardt et al.}{2005}]{bau05}
Baumgardt H., Grebel E. K., \& Kroupa P., 2005, MNRAS, 359, L1

\bibitem[\protect\citeauthoryear{Baumgardt \& Kroupa}{2007}]{Baumgardt07}
Baumgardt H., Kroupa P., 2007, MNRAS, 380, 1589

\bibitem[\protect\citeauthoryear{Baumgardt, De Marchi, \& Kroupa}{2008}]{Baumgardt08a}
Baumgardt H., De Marchi G., Kroupa P., 2008a, ApJ, 685, 247

\bibitem[\protect\citeauthoryear{Baumgardt, Kroupa, \& Parmentier}{2008}]{Baumgardt08b}
Baumgardt H., Kroupa P., Parmentier G., 2008b, MNRAS, 384, 1231

\bibitem[\protect\citeauthoryear{Bonnell \& Davies}{1998}]{bon98}
Bonnell I. A. \& Davies M. B., 1998, MNRAS, 295, 691

\bibitem[\protect\citeauthoryear{De Marchi et al.}{2007}]{dem07}
De Marchi G., Paresce F., Pulone L., 2007, ApJ, 656, L65

\bibitem[\protect\citeauthoryear{Dekker et al.}{2000}]{dek00}
Dekker H., D'Odorico S., Kaufer A., Delabre B., \& Kotzlowski H.,
2000, Proc. SPIE, 4008, 534

\bibitem[\protect\citeauthoryear{Djorgovski \& King}{1986}]{djo86}
Djorgovski S. \& King I. 1986, ApJ, 305, L61–L65.

\bibitem[\protect\citeauthoryear{Dotter et al.}{2007}]{dot07}
Dotter A., Chaboyer B., Jevremovic D., Baron E., Ferguson J.W.,
Sarajedini A., \& Anderson J., 2007, AJ, 134, 376

\bibitem[\protect\citeauthoryear{Dotter et al.}{2008}]{dot08}
Dotter A., Sarajedini A., \& Yang S.-C., 2008, AJ, 136, 1407

\bibitem[\protect\citeauthoryear{Elmegreen}{2004}]{elm04}
Elmegreen B.G., 2004, MNRAS, 354, 367

\bibitem[\protect\citeauthoryear{Giersz \& Heggie}{2009}]{gie09}
Giersz M. \& Heggie D. C., 2009, MNRAS, 395, 1173.

\bibitem[\protect\citeauthoryear{Goodman \& Hut}{1989}]{goo89}
Goodman J. \& Hut P., 1989, Nature, 339, 40.

\bibitem[\protect\citeauthoryear{Goodwin}{1997}]{goo97}
Goodwin S. P., 1997, MNRAS, 286, 669

\bibitem[\protect\citeauthoryear{Goodwin}{1998}]{goo98}
Goodwin S. P., 1998, MNRAS, 294, 47

\bibitem[\protect\citeauthoryear{Haghi et al.}{2009}]{hag09}
Haghi H., Baumgardt H., Kroupa P., Grebel E. K., Hilker M., Jordi K., 2009, MNRAS, 395, 1549

\bibitem[\protect\citeauthoryear{Harfst et al.}{2009}]{har09}
Harfst S., Portegies Zwart S., Stolte A., 2009 (arXiv:0911.3058)

\bibitem[\protect\citeauthoryear{Harris}{1996}]{har96}
Harris W. E., 1996, AJ, 112, 1487

\bibitem[\protect\citeauthoryear{Heggie \& Hut}{2003}]{heg03}
Heggie D. C. \& Hut P., 2003, The Gravitational Million-Body
Problem, Cambridge: Cambridge University Press.

\bibitem[\protect\citeauthoryear{Hilker}{2006}]{hilk06}
Hilker M., 2006, A\&A, 448, 171.

\bibitem[\protect\citeauthoryear{Hurley et al.}{2000}]{hur00}
Hurley J. R., Pols O. R., Tout C. A., 2000, MNRAS 315, 543

\bibitem[\protect\citeauthoryear{Hut et al.}{1992}]{hut92}
Hut P., et al., 1992, PASP, 104, 981

\bibitem[\protect\citeauthoryear{Jordi et al.}{2009}]{jor09}
Jordi K., Grebel, E. K., Hilker, M., Baumgardt, H., Frank, M.,
Kroupa, P., Haghi, H., Cote, P., Djorgovski, S. G., 2009, 137, Issue
6, pp. 4586-4596.

\bibitem[\protect\citeauthoryear{King}{1966}]{kin66}
King I. R., 1966, AJ, 71, 64

\bibitem[\protect\citeauthoryear{Kruijssen \& Mieske}{2009}]{krui09}
Kruijssen, J. M. D.; Mieske, S.,  2009 A\&A, 500, 785

\bibitem[\protect\citeauthoryear{Krumholz et al.}{2009}]{krum09}
Krumholz M. R., Klein R. I., McKee C. F., Offner S. S. R.,
Cunningham A. J., 2009, Science, 323, 754

\bibitem[\protect\citeauthoryear{Kroupa, Gilmore, \& Tout}{1991}]{kro91}
Kroupa P., Gilmore G., Tout C.A., 1991, MNRAS, 251, 293

\bibitem[\protect\citeauthoryear{Kroupa}{1995a}]{kro95a}
Kroupa P., 1995a, MNRAS, 277, 1491

\bibitem[\protect\citeauthoryear{Kroupa}{1995b}]{kro95b}
Kroupa P., 1995b, MNRAS, 277, 1507

\bibitem[\protect\citeauthoryear{Kroupa}{2008}]{kro08}
Kroupa P., 2008, 2008, The Cambridge N-Body Lectures, Lecture Notes
in Physics, 760, 181

\bibitem[\protect\citeauthoryear{Kroupa}{2001}]{kro01}
Kroupa P., 2001, MNRAS, 322, 231

\bibitem[\protect\citeauthoryear{Kroupa}{2002}]{kro02}
Kroupa P., 2002, Science, 295, 82

\bibitem[\protect\citeauthoryear{Kroupa \& Boily}{2002}]{kroboi02}
Kroupa P., Boily C. M., 2002, MNRAS, 336, 1188

\bibitem[\protect\citeauthoryear{K\"upper \& Kroupa}{2010}]{Kuepper10}
K\"upper A.H.W., Kroupa P., 2010, ApJ, 716, 776

\bibitem[\protect\citeauthoryear{Marks, Kroupa, \& Baumgardt}{2008}]{Marks08}
Marks M., Kroupa P., Baumgardt H., 2008, MNRAS, 386, 2047

\bibitem[\protect\citeauthoryear{Marks \& Kroupa}{2010}]{Marks10}
Marks M., Kroupa P., 2010, MNRAS, 406, 2000

\bibitem[\protect\citeauthoryear{Mouri \& Taniguchi}{2002}]{mou02}
Mouri H., Taniguchi Y., 2002, ApJ, 580, 844

\bibitem[\protect\citeauthoryear{Parmentier \&
Gilmore}{2007}]{par07} Parmentier G., Gilmore G., 2007, MNRAS, 377,
352

\bibitem[\protect\citeauthoryear{Parmentier et al.}{2008}]{par08}
Parmentier, G., Goodwin, S. P., Kroupa, P., Baumgardt, H., 2008,
ApJ, 678, 347

\bibitem[\protect\citeauthoryear{Parmentier \&
Kroupa}{2007}]{par10} Parmentier G., Kroupa P., 2010, accepted in
MNRAS

\bibitem[\protect\citeauthoryear{Salpeter}{1955}]{sal55}
Salpeter E. E., 1955, ApJ, 121, 161

\bibitem[\protect\citeauthoryear{Sarajedini}{1997}]{sar97}
Sarajedini A., 1997, AJ, 113, 682

\bibitem[\protect\citeauthoryear{\u{S}ubr et al.}{2008}]{sub08}
\u{S}ubr L., Kroupa P., Baumgardt H., 2008, MNRAS, 385, 1673.

\bibitem[\protect\citeauthoryear{Tan et al.}{2006}]{tan06}
Tan J. C., Krumholz M. R., McKee C. F., 2006, ApJ, 641, L121

\bibitem[\protect\citeauthoryear{Trenti et al.}{2010}]{tre10}
Trenti M., Vesperini E. \& Pasquato M., 2010, ApJ, 708, 1598

\bibitem[\protect\citeauthoryear{Vesperini \& Heggie}{1997}]{ves97}
Vesperini E. \& Heggie D. C., 1997, MNRAS, 289, 898

\bibitem[\protect\citeauthoryear{Vesperini et al.}{2009}]{ves09a}
Vesperini E., McMillan S., Portegies Zwart S., 2009a, Ap\&SS, 177

\bibitem[\protect\citeauthoryear{Vesperini et al.}{2009}]{ves09b}
Vesperini E., McMillan S., Portegies Zwart S., 2009b, ApJ, 698, 615

\bibitem[\protect\citeauthoryear{Vesperini}{2009}]{ves09c}
Vesperini E., 2010, Philosophical Transactions of the Royal Society
A: Mathematical, Physical and Engineering Sciences, 368, 829
(arXiv:0911.0793)

\bibitem[\protect\citeauthoryear{Vogt et al.}{1994}]{vog94}
Vogt S. S., et al., 1994, Proc. SPIE, 2198, 362



\end{thebibliography}
\end{document}